\documentclass[mathleft
]{an}
\usepackage{graphicx}
\usepackage{times}
\usepackage{sidecap}
\begin{document}

\Pagespan{789}{}
\Yearpublication{2006}%
\Yearsubmission{2005}%
\Month{11}%
\Volume{999}%
\Issue{88}%

\title{Newly identified YSO candidates towards the LDN 1188}

\author{G. Marton \inst{1} \thanks{Corresponding author:
  \email{marton.gabor@csfk.mta.hu}\newline},
E. Vereb\'elyi \inst{1},
Cs. Kiss \inst{1}
\and J. Smidla \inst{2}
}
\titlerunning{}
\authorrunning{G. Marton et al. - YSO candidates towards the LDN 1188}
\institute{Konkoly Observatory, Research Centre for Astronomy and Earth Sciences, Hungarian Academy of Sciences, Konkoly Thege 15-17, H-1121 Budapest, Hungary
\and Department of Computer Science and Systems Technology, University of Pannonia, Egyetem u. 10, H-8200 Veszpr\'em, Hungary}

\received{30 May 2005}
\accepted{11 Nov 2005}
\publonline{later}


\abstract{We present an analysis of Young Stellar Object (YSO) candidates towards the LDN 1188 molecular cloud. The YSO candidates were selected from the WISE all-sky catalogue, based on a statistical method. We found 601 candidates in the region, and classified them as Class I, Flat and Class II YSOs. Groups were identified and described with the Minimal Spanning Tree (MST) method. Previously identified molecular cores show evidence of ongoing star formation at different stages throughout the cloud complex.}

\maketitle

\section{Introduction}
LDN 1188 (Lynds, 1962) is a dark cloud complex located nearby the S140/L1204 star forming molecular complex towards the Cepheus region. It is part of a giant ring around the Cep OB2 association (Kun, M., Bal\'azs, L. G., T\'oth, I. 1987, \'Abrah\'am P., Bal\'azs L. G., Kun M., 2000). The derived distance of the cloud is 910 pc (\'Abrah\'am et al. 1995), based on  IRAS 100 $\mu$m optical depth, with an estimated total mass of $\sim$1800 M$_{\sun}$. Star forming activity of LDN 1188 has been investigated in several papers. K\"onyves et al. (2004) reported a detailed study of three YSOs by using 2MASS, MSX, IRAS and ISO data, constructing the spectral energy distribution of the objects. \'Abrah\'am et al. (1995) found candidates for YSOs by detecting $H\alpha$ emission objects and selected 6 IRAS sources as YSOs from the 25 infrared point sources located in the region. 

In a recent paper Vereb\'elyi et al. (this issue) summarised the results of the molecular line measurements in this dark cloud complex, including NH$_3$, CS and submm observations at the 1.2 mm continuum. These measurements revealed that there are various evolutionary phases present in parallel in the dark could complex. In this paper we investigate the YSO content of LDN 1188 using public WISE point source catalog data (Cutri et al. 2012), with the aim to characterise the young population and compare their distribution with that of the molecular material.

\section{WISE data and YSO selection method}
WISE (Wide-field Infrared Survey Explorer, Wright et al. 2010) is the most recent infrared survey satellite. The point source catalog (Cutri et al. 2012) built from the observations contains photometric data on 563921584 sources observed at 4 wavelengths, 3.4, 4.6, 12 and 22 $\mu$m. The achieved angular resolution was 6.1$^{\prime\prime}$, 6.4$^{\prime\prime}$, 6.5$^{\prime\prime}$, and 12.0$^{\prime\prime}$, respectively, which is $\sim$5 times better than that of IRAS at 12$\mu$m. Also, while IRAS could detect a source with a brightness of at least 0.5 Jy at 12 $\mu$m, the detection limit of WISE was lower than 0.01 Jy at this wavelength.

The detailed YSO selection method will be further discussed in our forthcoming paper (Marton et al. 2013, in prep.). Here we present only a short summary: the selection method itself is a statistical method. The Quadratic Discriminant Analysis aims to differentiate groups in multi dimensional data, regardless its nature. During the multi-step process, one needs to define a training sample, which is then used to "teach" the algorithm the quadratic boundaries between the different groups. These boundaries are then applied to the data, and a probability is associated with each object indicating the likelihood of belonging to one or the other group. We used SIMBAD and VizieR databases during the teaching phase, and applied the boundaries to those WISE detections that had a signal-to-noise ratio of $>$7 in each band.

\begin{SCfigure*}
\centering
\includegraphics[width=14cm]{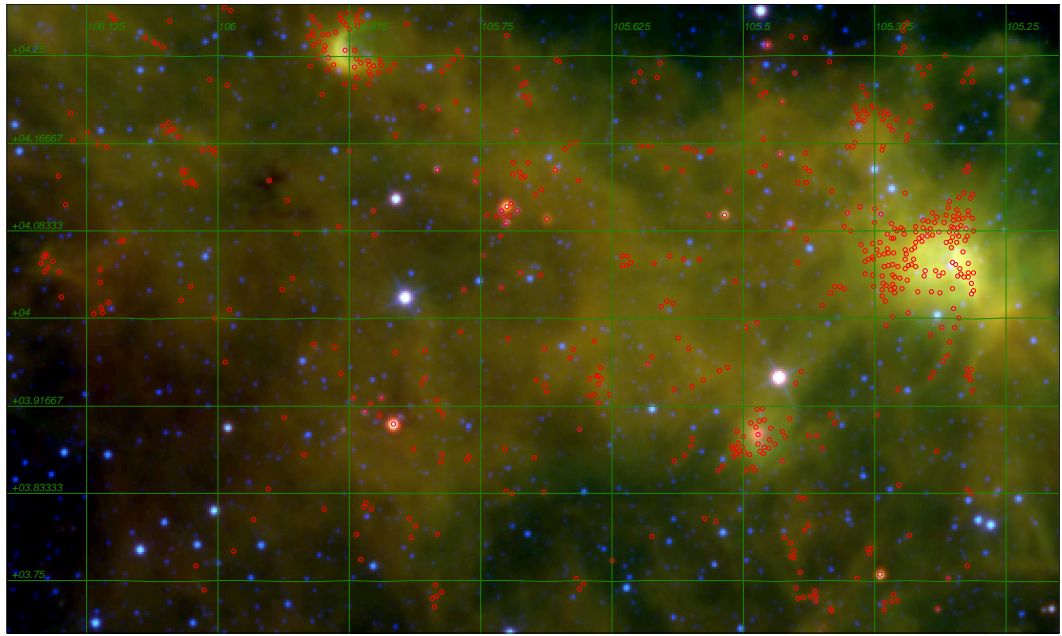}
\caption{Composite RGB image of the studied area around LDN 1188. Red, green and blue colours correspond to the WISE 4.6, 12 and 22 $\mu$m images, respectively. The selected YSO candidates are overlaid with small red circles. The green grid overlaid on the image represent the Galactic coordinate system.}
\label{fig1}
\end{SCfigure*}

\section{Results}
\subsection{YSO candidates in LDN 1188}
We found 601 YSO candidates in the studied area (105.2$<$l$<$106.2 and 3.7$<$b$<$4.3, see Fig. \ref{fig1}). Position correlation of the sources with the SIMBAD database shows that almost all sources are of previously unknown type. Only 7 counterparts have been found, when using the WISE point source coordinates as centres of a 5" radius search area. The source $[$ADM95$]$ 1 (IRAS 22129+6110 at $l$=105.29, $b$=+04.05) is located within a reflection nebula DG 180 and it is also close to the CO clump "A" (\'Abrah\'am et al. 1995). IRAS 22143+6120 (at $l$=105.52, $b$=+04.1) and 2MASS J22174025+6147025 (at $l$=105.79, $b$=+04.14) are known as single stars. CDS 1319 (at $l$=105.49, $b$=+03.89) is known as an emission line star closely located to the GN 22.15.0 reflection nebula. IRAS 22174+6121 (IRS-6 at $l$=105.83, $b$=+04.90, see Fig. \ref{fig3}) appears as an IR object in SIMBAD, while one source is closely located (at $l$=105.29, $b$=+04.05) to the GN 22.12.9 reflection nebula.

\subsection{YSO candidate classes}
In order to have an estimate on the evolutionary stages of all the candidates, we calculated the spectral indices ($\alpha$) for each object. The colour correction needed in the range of the potential temperature of these objects is relatively low ($\sim$1\%), therefore we did not apply this correction to the magnitudes. To calculate $\alpha$, we followed the method described in Majaess (2012), and used the formula: $\alpha$ $\approx 0.36(W_1-W_2)+ 0.58(W_2-W_3)+ 0.41(W_3-W_4)-2.90$, where W1 to W4 are the source magnitudes for the 4 WISE passbands from 3.6 to 22 $\mu$m, respectively.

The conventional criteria for the evolutionary stages features $\alpha>0.3, -0.3<\alpha<0.3, -1.6<\alpha<-0.3$ and $\alpha<-1.6$ for Class I, Flat, Class II and Class III objects, respectively. Applying these criteria to our calculated $\alpha$ values resulted in 50 sources classified as Class II objects, 180 Flat type objects and 371 Class I sources. No Class III objects were found. For the spatial distribution of the classes see Fig. \ref{fig4}.

\subsection{Group identification}

\begin{figure}
\vspace{-0.5cm}\hspace{-1cm}\includegraphics[width=9cm]{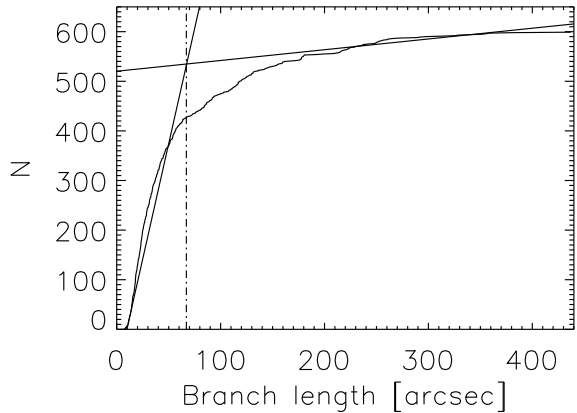}
\vspace{-0.3cm}
\caption{Cumulative distribution function of the MST branch length. Steep and shallow segments were approximated with single linear fits (solid lines). Their intersection defines the $L_{Cut}$ (dashed line).}
\label{fig2}
\end{figure}

We used the Minimal Spanning Tree (MST) method of Cartwright \& Whitworth (2004), as described in Gutermuth et al. (2009) and Beerer et al. (2010) to identify the groups. The construction of such a tree is described by Gower \& Ross (1969). Within the MST structure, groups can be separated as having ÒsmallÓ branch lengths between all members, i.e., less than some cutoff branch length ($L_{Cut}$).
The $L_{Cut}$ was calculated from the cumulative distribution function of the MST branch length, approximated by two linear segments (see Fig. \ref{fig2}). The shallow and the steep line segments connected at a branch length of 67 arcseconds, which value was then used to identify the individual groups within the MST network.

\begin{figure}
\hspace{1cm}\includegraphics[width=7cm]{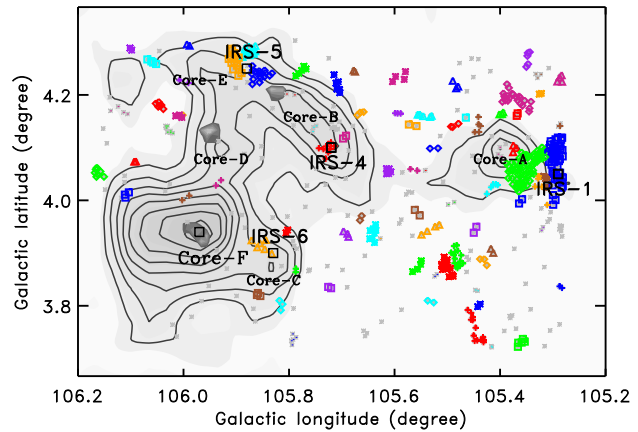}
\\
\\
\caption{Groups of YSO candidates identified with the MST method. Each group is presented with different colours and symbols. The identified number of groups with at least 2 members are 91. Candidates not associated with any group are plotted with gray asterisks. Sources are superimposed on the $^{13}$CO measurements of \'Abrah\'am et al. (1995, contour lines) and the NH$_3$ measurements of Vereb\'elyi et al. (this issue, filled grayscale contours). }
\label{fig3}
\end{figure}

\begin{figure}
\hspace{-0.5cm}\includegraphics[width=8.5cm]{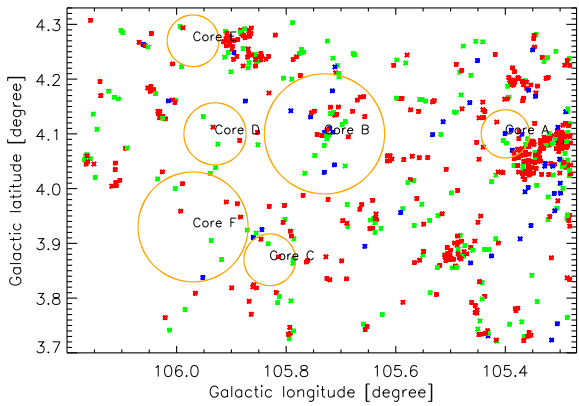}
\caption{YSO candidates and the molecular cloud cores of LDN 1188. YSO evolutionary stages are plotted as follows: red, green and blue asterisks present the Class I, Flat and Class II stages , respectively. $^{13}$CO cores from \'Abrah\'am et al. (1995) are plotted with solid orange circles.}
\label{fig4}
\end{figure}

In this way we were able to identify 91 groups within the network (see Fig. \ref{fig3}). The number of YSO candidates located in groups is 509 (85\% of all). The number of groups containing 4, or more candidates is 37. Table \ref{table2} list the properties of groups with at least 4 group embers. We investigated how changes in the linear approximation of the slopes affects the group identification. Fitting different sections of the distribution slightly changes the $L_{Cut}$. Also, 40\% increase in the $L_{Cut}$ resulted in only 15\% more groups, where the number of small groups was higher, but large ones started to merge. A 10\% lower $L_{Cut}$ increased the group number with only 5\%. In this case large groups tended to fall apart.

As Table \ref{table2} shows, the vast majority of the group members are Class II and Flat type objects. 262 from the 371 (71\%) Class I objects are bound within these large groups. 68 of them (18\%) are bound within small groups and 41 of them (11\%) are not bound in any groups. 
98 (54\%) of the 180 Flat type objects are within large groups, 48 (27\%) are within small groups and 34 (19\%) are not bound to any groups. 
27 (54\%) of the 50 Class II objects are within large groups, 7 (14\%) are within small groups and 16 (32\%) are not bound. 
Class I objects tend to be bound to large groups (having at least 4 group members), which are in their early evolutionary stages. The more evolved Flat and Class II objects are also mainly connected to larger groups, but they tend to lose their connection to those groups, since 19 and 32 percent of Flat and Class II objects are single sources in comparison to the 11\% of the Class I objects.

\section{Discussion}      

CO clumps were identified and analysed in \'Abrah\'am et al. (1995), and they also determined the size of the cloud cores. We used this apparent radius of the cores to determine if a star is associated with one of the cores. The distance limit for being in association with a core was set to 1.5 core radius and we also determined the average $\alpha$ value of the stars belonging to each CO core. A summary of the results and the core properties published in \'Abrah\'am et al. (1995) are presented in Table \ref{table4}. 

Core A: It is the core with the highest column density and with the highest YSO candidate count to M$_{\sun}$ ratio despite it's relatively low gas mass of 130 M$_{\odot}$. The average $\alpha$ value of the associated candidates was found to be the highest here among the cores, suggesting that most of the YSO candidates are very young, in Class I evolutionary stage. The two biggest groups with 73 and 52 members are also located in the vicinity. IRS-1 (IRAS 22129+6110) is a nearby source that might be associated with Core A. 

Core B: This is the core with the lowest column density in the studied area. The number of associated YSO candidates is the highest among the cores, but their average spectral index is close to that of the Flat class members, suggesting that these stars are relatively older. Only smaller groups of candidates can be found in the vicinity of the core. IRS-4 is apparently associated with the core.

Core C: The YSO content of the clump is similar to that of Core B with a smaller number of sources. The average spectral index is close to that of the Flat class. The column density of the core is higher than in the previous case, but the YSO candidate count to M$_{\sun}$ ratio is lower. The apparent position IRS-6 is close to the core with only small groups of YSOs. 

Core D: We found that Core D is associated with fewer number of candidates and the median $\alpha$ is lower than in the other cases, suggesting that it is a relatively old clump. Moreover, none of the associated candidates is a member of a group with at least 4 objects and the calculated number of YSO candidates over the core mass is the lowest among the cores. 

Core E: In the vicinity of this core we could identify several groups with $>$10 group members. The median of the estimated spectral index suggests that the majority of the sources are from Class I, in fact the second youngest population in LDN 1188. Although the infrared source IRS-5 is located near to the cloud, sub-millimetre observations could not identify any detectable source there (Vereb\'elyi et al., this issue).

Core F: The observed column density is the second highest among the cores, and this core has far the highest mass in the field. This core shows the lowest YSO candidate count to M$_{\sun}$ ratio in LDN 1188, in agreement with the results of Vereb\'elyi et al. (this issue) that this core is likely in an early evolutionary phase.

An overdensity of YSO candidates is observable in the direction of IRAS 22150+6109 (l=105.49 and b=3.89), which is known to be associated with an emission line OB star 3 + 60 9 (Wackerling 1970). 

\begin{table}
\centering
\caption{General properties of CO cloud cores and their estimated star formation activity. The columns are: core ID, galactic coordinates, median $\alpha$ value of the associated candidates, their standard deviation, number of the associated candidates, the number of stars over the core mass and the number density of the cores.
}
\label{table4}
\footnotesize\addtolength{\tabcolsep}{-5pt}
\begin{tabular}{|c|cc|c|c|c|c|c|}\hline
Clump & \multicolumn{2}{|c|}{Position} & $\alpha_{med}$ & $\sigma_{\alpha}$ & N$_{YSO}$ & N$_{YSO}$/M$_{cloud}$ & n\\
 & l$^{\circ}$ & b$^{\circ}$ & & & &M$_{\odot}^{-1}$& cm$^{-3}$\\
\hline
       A &        105.40 & 4.10 &         0.83 &            0.72 & 47 & 0.36 & 1770\\
       B &        105.73 & 4.10 &         0.31 &            0.64 & 66 & 0.25 & 600\\
       C &        105.83 & 3.87 &         0.30 &            0.66 & 19 & 0.12 & 930\\
       D &        105.93 & 4.10 &         -0.13 &            0.48 & 9 & 0.05 & 1120\\
       E &        105.97 & 4.27 &         0.54 &            0.71 & 15 & 0.17 & 1040\\
       F &        105.97 & 3.93 &         0.43 &            0.43 & 12 & 0.012 & 1140\\
\hline
\end{tabular}
\end{table}

\section{Summary}
601 YSO candidates were identified towards the LDN 1188 infrared dark cloud. They are located in 91 groups, from which 37 have 4 or more members. 371, 180 and 50 sources were classified as Class II, Flat and Class I sources, respectively, based on their estimated $\alpha$ spectral indices. Our results are in agreement with the findings of the molecular material survey by Vereb\'elyi et al. (this issue) concerning the evolutionary stages of the different parts of this cloud complex.

\begin{table}
\centering
\caption{Properties of groups with at least 4 group members. Columns are as follows: The group's ID number, the galactic coordinates of the group mass centre, and the number of Class I, Flat and Class II stars within the group.}
\label{table2}
\footnotesize\addtolength{\tabcolsep}{-2pt}
\begin{tabular}{|c|c|c|c|c|c|}\hline
ID & l & b & Class I & Flat & Class II \\
\hline
       1 &        105.45012 &        3.7782540 &            9 &            3 &            1\\
       3 &        105.44245 &        3.8283770 &            2 &            2 &        0\\
       8 &        105.50598 &        3.8776820 &           10 &            5 &        0\\
       9 &        105.55706 &        3.9274430 &           14 &            2 &        0\\
      12 &        105.55048 &        3.9040500 &            2 &            2 &        0\\
      16 &        105.61503 &        3.9005360 &            3 &            1 &        0\\
      19 &        105.58165 &        4.1385340 &            2 &            3 &        0\\
      23 &        105.78979 &        4.2000680 &            2 &            2 &        0\\
      29 &        105.40819 &        4.0657340 &            5 &            1 &        0\\
      30 &        105.57775 &        4.0113840 &            5 &        0 &        0\\
      32 &        105.35107 &        4.0568050 &           60 &           12 &            1\\
      34 &        105.32864 &        4.0804650 &           32 &           12 &            8\\
      35 &        105.32179 &        4.0612100 &            1 &            2 &            2\\
      36 &        105.35825 &        4.0890170 &            2 &        0 &            3\\
      37 &        105.35757 &        4.1067460 &            2 &            1 &            1\\
      39 &        105.32298 &        4.0994040 &           18 &            2 &            2\\
      42 &        105.39787 &        3.9452450 &            4 &            2 &        0\\
      43 &        105.37060 &        4.0606210 &            3 &            2 &        0\\
      46 &        105.34638 &        4.1294340 &            3 &            1 &            1\\
      47 &        105.36664 &        4.1650330 &            3 &            1 &        0\\
      48 &        105.36978 &        3.9314960 &            4 &            1 &        0\\
      51 &        105.53272 &        3.9368650 &            5 &            4 &        0\\
      56 &        105.67760 &        4.1627060 &            4 &            2 &        0\\
      57 &        105.72053 &        4.1784420 &            3 &            4 &            3\\
      58 &        105.54562 &        4.1308120 &            1 &            4 &            1\\
      60 &        105.70778 &        4.1616880 &            3 &            1 &        0\\
      66 &        105.78670 &        3.8376040 &            3 &            2 &        0\\
      68 &        105.86796 &        4.1670290 &            1 &            3 &            2\\
      72 &        105.57941 &        4.0978430 &            2 &            3 &        0\\
      74 &        105.83225 &        4.0858290 &           13 &            4 &        0\\
      75 &        105.90989 &        4.2595540 &           10 &            2 &        0\\
      76 &        105.90340 &        4.2405880 &           13 &            2 &            1\\
      80 &        105.87865 &        4.2042930 &            6 &            1 &        0\\
      81 &        106.04801 &        4.1799080 &            5 &            1 &        0\\
      83 &        105.87592 &        4.2697040 &            2 &            1 &            1\\
      87 &        106.02504 &        4.2555210 &            1 &            3 &        0\\
      88 &        106.02050 &        4.1274220 &            4 &            4 &        0\\
\hline
\end{tabular}
\end{table}      
      
\acknowledgements
Valuable comments and corrections of our anonymus referee are acknowledged. This work has been supported by the following grants: i) PECS contract no. 98073 of the Hungarian Space Office and the European Space Agency, ii) Hungarian Research Fund (OTKA) grants nr. 101939 and 104607. This research has made use of the SIMBAD database, operated at CDS, Strasbourg, France. This publication makes use of data products from the Wide-field Infrared Survey Explorer, which is a joint project of the University of California, Los Angeles, and the Jet Propulsion Laboratory/California Institute of Technology, funded by the National Aeronautics and Space Administration.

\end{document}